\documentstyle[12pt]{article}

\newcommand{\sect}[1]{\setcounter{equation}{0}\section{#1}}

\newskip\humongous \humongous=0pt plus 1000pt minus 1000pt
\def\caja{\mathsurround=0pt}
\def\eqalign#1{\,\vcenter{\openup2\jot \caja
        \ialign{\strut \hfil$\displaystyle{##}$&$
        \displaystyle{{}##}$\hfil\crcr#1\crcr}}\,}
\newif\ifdtup

\def\bseq{\begin{subequation}}  
\def\eseq{\end{subequation}}
\def\bsea{\begin{subeqnarray}}  
\def\esea{\end{subeqnarray}}

\def\de{\nabla}
  
\def\pa{\partial}
\def\cdm{{\cal D}_{\mm}}
\def\cdp{{\cal D}_{\pp}}

\def\Bar#1{\overline{#1}}                       
\def\lin{\vrule width0.5pt height5pt depth1pt}
\def\dpx{{{ =\hskip-3.75pt{\lin}}\hskip3.75pt }}

\def\Tilde#1{\widetilde{#1}}                    
\def\un#1{\underline {#1}}

\def\half{{\scriptstyle{1 \over 2}}}
\def\sint{{\int d^2\sigma d\zeta^-}}

\def\ha{\frac12}
\def\leftrightarrowfill{$\mathsurround=0pt \mathord\leftarrow \mkern-6mu
        \cleaders\hbox{$\mkern-2mu \mathord- \mkern-2mu$}\hfill
        \mkern-6mu \mathord\rightarrow$}
\def\dvec#1{\vbox{\ialign{##\crcr
        \leftrightarrowfill\crcr\noalign{\kern-1pt\nointerlineskip}
        $\hfil\displaystyle{#1}\hfil$\crcr}}}           
\def\bo{{\raise.15ex\hbox{\large$\Box$}}}

\newcommand{\beq}{\begin{equation}}
\newcommand{\eeq}{\end{equation}}
\newcommand{\bea}{\begin{eqnarray}}
\newcommand{\eea}{\end{eqnarray}}
\newcommand{\ena}{\end{eqnarray}}

\renewcommand{\a}{\alpha}
\renewcommand{\b}{\beta}

\renewcommand{\d}{\delta}

\newcommand{\G}{\Gamma}

\newcommand{\z}{\zeta}

\renewcommand{\l}{\lambda}
\renewcommand{\L}{\Lambda}
\newcommand{\m}{\mu}

\newcommand{\F}{\Phi}

\renewcommand{\P}{\Pi}

\newcommand{\s}{\sigma}
\renewcommand{\S}{\Sigma}
\renewcommand{\t}{\tau}

\def\Mb{\kern 2pt\mathchoice
            {
             \vbox{\hrule width10pt height 0.4pt depth 0pt
                 \kern 1.2pt\hbox{\kern -2pt$\displaystyle M$}}}
            {
                 \vbox{\hrule width10pt height 0.4pt depth 0pt
                 \kern 1.2pt\hbox{\kern -2pt$\textstyle M$}}}
            {
\vbox{\hrule width6pt height 0.4pt depth 0pt
                 \kern 1.0pt\hbox{\kern -2pt$\scriptstyle M$}}}
            {
                 \vbox{\hrule width5pt height 0.4pt depth 0pt
                 \kern 0.8pt\hbox{\kern -2pt$\scriptscriptstyle M$}}}}

\def\Sb{\kern 2pt\mathchoice
            {
                 \vbox{\hrule width6pt height 0.4pt depth 0pt
                 \kern 1.2pt\hbox{\kern -2pt$\displaystyle S$}}}
            {
                 \vbox{\hrule width6pt height 0.4pt depth 0pt
                 \kern 1.2pt\hbox{\kern -2pt$\textstyle S$}}}
            {
                 \vbox{\hrule width3.5pt height 0.4pt depth 0pt
                 \kern 1.0pt\hbox{\kern -2pt$\scriptstyle S$}}}
            {
                 \vbox{\hrule width3pt height 0.4pt depth 0pt
                 \kern 0.8pt\hbox{\kern -2pt$\scriptscriptstyle S$}}}}

\def\Rb{\kern 2pt\mathchoice
            {
                 \vbox{\hrule width5.5pt height 0.4pt depth 0pt
                 \kern 1.2pt\hbox{\kern -2.5pt$\displaystyle R$}}}
            {
                 \vbox{\hrule width5.5pt height 0.4pt depth 0pt
                 \kern 1.2pt\hbox{\kern -2.5pt$\textstyle R$}}}
            {
                 \vbox{\hrule width3.5pt height 0.4pt depth 0pt
                 \kern 1.0pt\hbox{\kern -2.2pt$\scriptstyle R$}}}
            {
                 \vbox{\hrule width3pt height 0.4pt depth 0pt
                 \kern 0.8pt\hbox{\kern -2.2pt$\scriptscriptstyle R$}}}}

  \def\pp{{\mathchoice
          {
              \kern 1pt%
              \raise 1pt
              \vbox{\hrule width5pt height0.4pt depth0pt
                    \kern -2pt
                    \hbox{\kern 2.3pt
                          \vrule width0.4pt height6pt depth0pt
                          }
                    \kern -2pt
                    \hrule width5pt height0.4pt depth0pt}%
                    \kern 1pt
           }
            {
              \kern 1pt%
              \raise 1pt
              \vbox{\hrule width4.3pt height0.4pt depth0pt
                    \kern -1.8pt
                    \hbox{\kern 1.95pt
                          \vrule width0.4pt height5.4pt depth0pt
                          }
                    \kern -1.8pt
                    \hrule width4.3pt height0.4pt depth0pt}%
                    \kern 1pt
            }
            {
              \kern 0.5pt%
              \raise 1pt
              \vbox{\hrule width4.0pt height0.3pt depth0pt
                    \kern -1.9pt  
                    \hbox{\kern 1.85pt
                          \vrule width0.3pt height5.7pt depth0pt
                          }
                    \kern -1.9pt
                    \hrule width4.0pt height0.3pt depth0pt}%
                    \kern 0.5pt
            }
            {
              \kern 0.5pt%
              \raise 1pt
              \vbox{\hrule width3.6pt height0.3pt depth0pt
                    \kern -1.5pt
                    \hbox{\kern 1.65pt
                          \vrule width0.3pt height4.5pt depth0pt
                          }
                    \kern -1.5pt
                    \hrule width3.6pt height0.3pt depth0pt}%
                    \kern 0.5pt
            }
        }}

  \def\mm{{\mathchoice
                       {
                             \kern 1pt
               \raise 1pt    \vbox{\hrule width5pt height0.4pt depth0pt
                                  \kern 2pt
                                  \hrule width5pt height0.4pt depth0pt}
                             \kern 1pt}
                       {
                            \kern 1pt
               \raise 1pt \vbox{\hrule width4.3pt height0.4pt depth0pt
                                  \kern 1.8pt
                                  \hrule width4.3pt height0.4pt depth0pt}
                             \kern 1pt}
                       {
                            \kern 0.5pt
               \raise 1pt
                            \vbox{\hrule width4.0pt height0.3pt depth0pt
                                  \kern 1.9pt
                                  \hrule width4.0pt height0.3pt depth0pt}
                            \kern 1pt}
                       {
                           \kern 0.5pt
             \raise 1pt  \vbox{\hrule width3.6pt height0.3pt depth0pt
                                  \kern 1.5pt
                                  \hrule width3.6pt height0.3pt depth0pt}
                           \kern 0.5pt}
                       }}

\def\pd{{\kern0.5pt
                   + \kern-5.05pt \raise5.8pt\hbox{$\textstyle.$}\kern
0.5pt}}

\def\pmd{{\kern0.5pt
                  \pm \kern-5.05pt \raise6.3pt\hbox{$\textstyle.$}\kern1.5pt}}

\def\md{{\mathchoice
   {
      {{\kern 1pt - \kern-6.2pt \raise5pt\hbox{$\textstyle.$}\kern 1pt}}}
    {
      {{\kern 1pt - \kern-6.2pt \raise5pt\hbox{$\textstyle.$}\kern 1pt}}}
    {
      {\kern0.5pt - \kern-5.05pt \raise3.4pt\hbox{$\textstyle.$}\kern0.5pt}}
    {
      {\kern0.5pt - \kern-5.05pt \raise3.4pt\hbox{$\textstyle.$}\kern0.5pt}}}}

\def\Sc{\scriptstyle}

\def\Mb{\kern 2pt\mathchoice
            {
             \vbox{\hrule width10pt height 0.4pt depth 0pt
                 \kern 1.2pt\hbox{\kern -2pt$\displaystyle M$}}}
            {
                 \vbox{\hrule width10pt height 0.4pt depth 0pt
                 \kern 1.2pt\hbox{\kern -2pt$\textstyle M$}}}
            {
\vbox{\hrule width6pt height 0.4pt depth 0pt
                 \kern 1.0pt\hbox{\kern -2pt$\scriptstyle M$}}}
            {
                 \vbox{\hrule width5pt height 0.4pt depth 0pt
                 \kern 0.8pt\hbox{\kern -2pt$\scriptscriptstyle M$}}}}

\def\Sb{\kern 2pt\mathchoice
            {
                 \vbox{\hrule width6pt height 0.4pt depth 0pt
                 \kern 1.2pt\hbox{\kern -2pt$\displaystyle S$}}}
            {
                 \vbox{\hrule width6pt height 0.4pt depth 0pt
                 \kern 1.2pt\hbox{\kern -2pt$\textstyle S$}}}
            {
                 \vbox{\hrule width3.5pt height 0.4pt depth 0pt
                 \kern 1.0pt\hbox{\kern -2pt$\scriptstyle S$}}}
            {
                 \vbox{\hrule width3pt height 0.4pt depth 0pt
                 \kern 0.8pt\hbox{\kern -2pt$\scriptscriptstyle S$}}}}

\def\Rb{\kern 2pt\mathchoice
            {
                 \vbox{\hrule width5.5pt height 0.4pt depth 0pt
                 \kern 1.2pt\hbox{\kern -2.5pt$\displaystyle R$}}}
            {
                 \vbox{\hrule width5.5pt height 0.4pt depth 0pt
                 \kern 1.2pt\hbox{\kern -2.5pt$\textstyle R$}}}
            {
                 \vbox{\hrule width3.5pt height 0.4pt depth 0pt
                 \kern 1.0pt\hbox{\kern -2.2pt$\scriptstyle R$}}}
            {
                 \vbox{\hrule width3pt height 0.4pt depth 0pt
                 \kern 0.8pt\hbox{\kern -2.2pt$\scriptscriptstyle R$}}}}

\def\headpic{           
  \indent
  \setlength{\unitlength}{.4mm}
  \thinlines
  \par
  \begin{picture}(29,16)
  \put(165,16){\line(1,0){4}}
  \put(170,16){\line(1,0){4}}
  \put(180,16){\line(1,0){4}}
  \put(175,0){\line(1,0){4}}
  \put(180,0){\line(1,0){4}}
  \put(185,0){\line(1,0){4}}
  \put(169,0){\line(0,1){16}}
  \put(170,0){\line(0,1){16}}
  \put(179,0){\line(0,1){16}}
  \put(180,0){\line(0,1){16}}
  \put(184,0){\line(0,1){16}}
  \put(185,0){\line(0,1){16}}
  \put(169,16){\oval(8,32)[bl]}
  \put(170,16){\oval(8,32)[br]}
  \put(179,0){\oval(8,32)[tl]}
  \put(185,0){\oval(8,32)[tr]}
  \end{picture}
  \par\vskip-6.5mm
  \thicklines}

\def\border{           
  \setlength{\unitlength}{1mm}
  \newcount\xco
  \newcount\yco
  \xco=-21
  \yco=12
  \begin{picture}(140,0)
  \put(\xco,\yco){$\ktl$}
  \advance\yco by-1
  {\loop
  \put(\xco,\yco){$\kcr$}
  \advance\yco by-2
  \ifnum\yco>-240
  \repeat
  \put(\xco,\yco){$\kbl$}}
  \xco=158
  \yco=12
  \put(\xco,\yco){$\ktr$}
  \advance\yco by-1
  {\loop
  \put(\xco,\yco){$\kcr$}
  \advance\yco by-2
  \ifnum\yco>-240
  \repeat
  \put(\xco,\yco){$\kbr$}}
  \put(-20,13){\tiny University of Maryland Elementary Particle
Physics University of Maryland Elementary Particle Physics University of
Maryland Elementary Particle Physics}
  \put(-20,-241.5){\tiny University of Maryland Elementary
Particle Physics University of Maryland Elementary Particle Physics
University of Maryland Elementary Particle Physics}
  \end{picture}
  \par\vskip-8mm}

\def\Sc#1{{\hbox{\sc #1}}}  
\font\ro=cmsy10       
\def\kcr{{\hbox{\ro \char'170}}}    
\def\ktl{{\hbox{\ro \char'170}}}  
\def\ktr{{\hbox{\ro \char'170}}}  
\def\kbl{{\hbox{\ro \char'170}}}  
\def\kbr{{\hbox{\ro \char'170}}}  

\def\endtitle{\end{quotation}\newpage}     

\begin{document}

\def\gfrac#1#2{\frac {\scriptstyle{#1}}
  {\mbox{\raisebox{-.6ex}{$\scriptstyle{#2}$}}}}
\def\gg{{\hbox{\sc g}}}
\border\headpic {\hbox to\hsize{April 1997 \hfill {UMDEPP 97-104}}}
\par
\setlength{\oddsidemargin}{0.3in}
\setlength{\evensidemargin}{-0.3in}
\begin{center}
\vglue .04in
{\large\bf Type -B/ -O Bosonic String Sigma-Models}
\\[.6in]

S. James Gates, Jr.\footnote{gates@umdhep.umd.edu}   \\[.1in]
{\it Department of Physics\\
University of Maryland at College Park\\
College Park, MD 20742-4111, USA\\
${~~}$\\
{\rm {and}}\\
${~~}$\\
{\rm {V.G.J. Rodgers}}\footnote{vincent@hepaxp.physics.uiowa.edu}\\
Department of Physics and Astronomy\\
University of Iowa\\
Iowa City, Iowa~~52242--1479} 
\\ [1.8in]

{\bf ABSTRACT}\\[.002in]
\end{center}
\begin{quotation}
{We provide world sheet non-supersymmetrical actions that describe the 
coupling of a bosonic string to the tachyon and massless states of  
both the type-B and type-O theories.  The type-B theory is derived as a 
truncation and chiral doubling of the Ramond-Ramond sector in our previous 
model that connected the (1,0) heterotic string to a 10D, type-IIB 
supergravity background.   The type-O theory then follows from a  
``fermionization'' of the type-B theory.}
\endtitle

\sect{Introduction}

~~~~Approximately two years ago, we pointed out a curious feature \cite{Idea} 
of (1,0) supergravity NSR non-linear $\s$-models and type-IIB supergravity.
Although totally unexpected at the time, we formulated a coupling of a 
heterotic string \cite{HET} to a background that consisted 
of the massless bosonic fields of the type-IIB supergravity theory.  The
reason for this to be unexpected was that it had always been thought that 
the type-IIB supergravity theory (whose superspace description is well 
known \cite{Typ2BSG}) was only associated with the type-IIB superstring 
of Green and Schwarz \cite{GS2}.  From this point of view it was very 
``unnatural'' to find a heterotic model associated with the massless 
bosonic fields of type-IIB supergravity.  

On the other hand, the construction of a (1,0) NSR non-linear $\s$-model 
coupled to a type-IIB supergravity background cleared up an earlier puzzle.   
Several years prior to the work in \cite{Idea}, it was noted that the (1,0) 
heterotic string nonlinear $\s$-model possessed exactly the right properties 
to allow a coupling to a 4D, N = 8 supergravity background also \cite{GN}.  
This 4D result can now be viewed as the simple dimensional reduction of 
the type-IIB heterotic string nonlinear $\s$-model.

In order to accomplish the result in \cite{Idea}, we introduced 
a feature which had not appeared in the context of stringy non-linear 
$\s$-models prior to that time.  The idea was that when certain combinations
of p-forms occur in the Ramond-Ramond sector of a superstring theory, they 
correspond to the introduction of WZNW models on the world sheet of the 
superstring where the currents of the WZNW model do not represent internal 
symmetries but instead are associated with the Clifford algebra of the the 
target spacetime.  This concept has reappeared recently in discussions of 
Dirichlet p-branes and type-IIB supergravity \cite{Nil}.  

Since the model of \cite{Idea} is a (1,0) theory, when analyzed in terms of 
its left and right handed degrees of freedom, one finds the expected 
result that the left-handed sector is supersymmetric while the 
right-handed sector is non-supersymmetric.  This observation 
is the key to the construction of the non-supersymmetric nonlinear
$\s$-model for the type-B string.  In the language of 2D field theory
we well know how to take an ordinary boson and separate it into
its left-handed and right-handed segments by use of chiral bosons
\cite{Sieg}.  Once this is done, the original left-handed supersymmetric
sector may be disgarded.  However, in order to have a consistent
closed bosonic string, it is required to have some non-supersymmetric
left-handed sector.  A new left-handed sector may be introduced
by constructing the ``mirror'' of the right-handed sector of the
original (1,0) type-IIB heterotic string non-linear $\s$-model.

There are two steps required for constructing the mirror. First,
all the 2D fields that depended only the $\s^{\mm}$-coordinate 
on-shell must be replaced by fields that depend on $\s^{\pp}$-coordinate 
on-shell\footnote{In the language of conformal field theory this 
corresponds to $X(z) \to {\Tilde X} ({\bar z})$.}. Also the target 
spacetime Clifford algebra of the original Ramond-Ramond sector of 
the (1,0) model must be replaced by its target spacetime chiral 
reflection\footnote{All the undotted 10D spinor indices must be 
replaced by dotted 10D spinor indices.}.  This mirror is the `glued' 
back to the original right handed sector and the result is the type-B
nonlinear $\s$-model with precisely the couplings to the massless 
spectrum of the type-B string.

\sect{Review of NSR Supergravity-Heterotic Sigma-Models in (1,0) Superspace}

~~~~Although the original formulation of the heterotic string \cite{HET}
was {\it {not}} as a (1.0) superfield theory, with the work of \cite{BGM}, 
it became possible to show that the action given by
\begin{equation}
S_{HET-1} ~=~ { 1 \over 4 \pi {\a}' } \sint E^{-1} \Big\{ ~   i  
\eta_{\un m \un n} ~ (\nabla_+ { X}^{\un m}) \, (\nabla_{\mm} 
{X}^{\un n}) \, - \,  (\, \eta_-{}^{\hat I } \nabla_+ \eta_-{}^{\hat 
I} \,) ~  \Big\} ~~~,
\end{equation}
upon gauge-fixing to eliminate the (1,0) supergravity fields yields
the action of Gross et.~al. in the fermionic formulation (i.e. ${\hat I}
= 1, ..., 32$).  A second formulation (also given by Gross et.~al.) 
utilizing chiral bosons \cite{Sieg}  whose (1,0) superspace form is
\begin{equation}
\eqalign{
S_{HET-2} ~=~ \sint &E^{-1} \Big\{ \, { 1 \over 4 \pi {\a}' } [ \,  i  
\eta_{\un m \un n} ~ (\nabla_+ { X}^{\un m}) \,  (\nabla_{\mm} {\bf 
X}^{\un n}) \, ]  ~+~ {~~~~~~~~~~~~~~~~~~} \cr
&{~~~~~~~~} i {1 \over 2 }[\, (\nabla_+ {\Phi_{R}}^{ \hat a } )\,( 
\nabla_{\mm}{\Phi_{R}}^{ \hat a }) ~+~ \Lambda_{\mm}{}^{\pp} ( \nabla_+ 
{\Phi_{R}}^{\hat a} ) \, ( \nabla_{\pp}{\Phi_{R}}^{\hat a })~ ] 
\,  \Big\} ~~~, }  
\end{equation}
where ${\hat a} = 1, ...16$ was also described.  Neither of these two 
formulations  allows for a manifest realization of the $E_8 \otimes E_8$ 
symmetry of heterotic string theory.  To accomplish this, it is necessary 
to carry out a non-abelian bosonization \cite{GaSg} of the fermionic 
superfields ($\eta_- {}^{\hat I}$) in the first action above\footnote{The 
action in (2.2) may be regarded as an abelian bozonization of (2.1).}. 
\begin{equation}
\eqalign{
S_{HET-3} ~=~ { 1 \over 4 \pi {\a}' } &\sint E^{-1} [ \,  i  \eta_{\un m 
\un n} ~ (\nabla_+ { X}^{\un m}) \,  (\nabla_{\mm} {X}^{\un n}) 
\, ] {~~~~~~~~~~~~~~~~~~}\cr
-{1 \over 2 \pi } &\sint E^{-1}  i \half Tr \{ ~ R_+ R_{=} ~+ i  
\L_{=}{}^{\dpx} R_+ \de_+ R_+ \cr 
&{~~~~~~~~~~~~~~~~~~~~~~~~~} ~+ {\scriptstyle{2\over3}} \L_{=}{}^{\dpx} 
\{ \, R_+  \,  , \, R_+ \, \} ~ R_+  \cr
~+  &\int_{0}^1 d \, y ~[ \, ( {d \widetilde U \over dy} \widetilde 
       U^{-1} \, ) [ \, \de_{=}((\de_+\widetilde U ) \widetilde U^{-1})
        ~-~  \de_{+}((\de_{=}\widetilde U ) \widetilde U^{-1} ) \, 
]~  \}  ~~~. ~~~~~~} 
\end{equation}
where the following definitions are used,
\begin{equation}
R_a ~\equiv~ U^{-1}  \de_a U ~~~,~~~ U  ~ \equiv ~ exp\Big[ \, i
{\Phi_{R}}^{\hat a } t_{\hat a } \Big] ~~~. 
\end{equation}
In this last equation, the quantities ${\Phi_{R}}^{\hat a }$ constitute
496 righton (1,0) superfields containing the same number of component 
rightons.  From this last equation it is clear that in order to give 
a complete specification of the action in (2.3), it is necessary to 
introduce a set of matrices (above denoted by $t_{\hat a }$) that form 
a Lie algebra. In fact, the action of (2.3) permits us to associate a
{\it {distinct}} world sheet action with {\it {each}} consistent 10D 
heterotic string.  It has long been known \cite{het10s} that the 
groups that lead to tachyon-free 10D heterotic strings are; $E_8 
\otimes E_8$, $SO(32)$ and $SO(16) \otimes SO(16)$.  Thus the matrices 
$t_{\hat a }$ may be chosen to provide representations of these algebras 
and there is no need to introduce winding modes to construct the 
$E_8 \otimes E_8$ representation.

No background fields at all appear in the actions above. The way to introduce
background spacetime fields is of course well known \cite{cups}. The NS-NS 
bosonic fields ($g_{ \un m \un n}$, $b_{ \un m \un n} $ and $\Phi$) are made 
to appear by replacing the first line of (2.3) by 
\begin{equation}
\eqalign{
S_{NS} &=~ { 1 \over 2 \pi {\a}' } \sint E^{-1} [  i \half ( g_{ \un m 
\un n}(X) ~ + ~ b_{ \un m \un n}(X) ) ~ (\nabla_+ { X}^{\un m}) 
(\nabla_{=} { X}^{\un n}) ~ ]  \cr
&{~~~~~~~}+~ \sint E^{-1} \Phi(X) {\S}^+  \cr
&\equiv~ \int d^2 \s d\z^- E^{-1}
 [i \frac12 (\eta_{\un a \un b} + b_{\un a \un b}(X))
\P_{+} {}^{\un a} \P_{=} {}^{\un b}
~+~  \Phi(X) \S^+ ~ ] ~~, }
\end{equation}
where $\P_{+} {}^{\un a} \equiv ( 1/ \sqrt{ 2 \pi {\a}'})(  \nabla_+ { 
X}^{\un m}) e_{\un m} {}^{\un a} $ and
$\P_{=} {}^{\un a} \equiv ( 1/ \sqrt{ 2 \pi {\a}'})(  \nabla_= { 
X}^{\un m}) e_{\un m} {}^{\un a} $.

The introduction of the R-R bosonic fields (i.e. the spacetime gauge 
fields $A_{ \un a} {}_{ \hat I \hat J }$) for the internal symmetry 
groups can be carried out for the $SO(32)$ and $SO(16) \otimes SO(16)$ 
theories quite easily \cite{SEN}. The last term in (2.1) is replaced by, 
\begin{equation}
S_{R} ~=~  - \half \, \sint E^{-1}[~ \eta_-{}^{\hat I } \nabla_+ 
\eta_-{}^{\hat I} ~+~  \P_{+} {}^{\un a} \, \eta_-{}^{\hat I } A_{\un a}
{}_{ \hat I \hat J }(X) \eta_-{}^{\hat J } ~ ] ~~~.
\end{equation}
However, the introduction of the R-R bosonic fields in the $E_8 
\otimes E_8$ theory can {\it {only}} be done by modifying the
last three terms in (2.3). These must be replaced by \cite{BDG},
\begin{equation}
\eqalign{
{S'}_R ~=~ - {1 \over 2 \pi }
&\sint E^{-1}  i \half Tr \{ ~ ( R_+ ~+~ 2 \G_+)  R_{=} \cr
& ~~+ i  \L_{=}{}^{\dpx} (R_+ ~+~ \G_+ ) \de_+ (R_+ ~+~ \G_+)  \cr
&~~+ {2\over3} \L_{=}{}^{\dpx} \{ \, (R_+ ~+~ \G_+) \, , ~
        (R_+ ~+~ \G_+) \, \} ~ (R_+ ~-~ \ha \G_+) \cr
&~~+  \int_{0}^1 d \, y ~[ \, ( {d \widetilde U \over dy} \widetilde 
       U^{-1} \, ) [ \, \de_{=}((\de_+\widetilde U ) \widetilde U^{-1})
        ~-~  \de_{+}((\de_{=}\widetilde U ) \widetilde U^{-1} ) \, 
]~  \}  ~~~. } 
\end{equation}
where $\G_+  ~ \equiv ~ \P_{+} {}^{\un a} A_{\un a} {}^{\hat a} 
(X) t_{\hat a}$. Since the non-abelian bosonized theory offers the
most complete description, we will only utilize it in the subsequent
discussion.

In addition to these three 10D tachyon-free heterotic string theories,
there are also a number of other non-supersymmetric theories that contain 
a tachyon \cite{het10s}.  These include $SO(32)$, $E_8 \otimes SO(16)$, $SO(24)
\otimes SO(8)$, $(E_7 \otimes SU(2))^2$, $SU(16) \otimes U(1)$,
$E_8$. Although it appears not generally known, the (1,0) supergeometry
also permits the introduction of a coupling of the tachyon to
the worldsheet of the heterotic string. This is accomplished
by a slight generalization of a result presented some time
ago \cite{BGM}. The construction requires the introduction of
a world sheet minus spinor superfield $\Psi_-$ that appears
in the action
\begin{equation}
S_{tachyon} ~=~ \sint E^{-1} [~  - \half (\, \Psi_-
             \nabla_+ \Psi_- \,)  ~+~ i T(X) \Psi_- ~]  ~~~.
\end{equation}
In the limit where $T(X) = 1$, this action introduces a cosmological 
constant on the worldsheet. Note that each of the non-supersymmetric
models corresponds to a distinct choice of the matrix generators
$t_{\hat a}$ for the groups listed above.

Up until our work of \cite{Idea}, as far as was known, the models
above described all 10D (1,0) supergravity NSR heterotic $\s$-models.
With this work we proposed the addition of one more such model.
The introduction of this final model, begins with the observation that
the $E_8 \otimes E_8$ algebra in (2.4) may be replaced by another 
(non-compact) Lie algebra so that the new group elements take the 
form
\begin{equation}
U \, \to\, U_{R-1} ~\equiv~ exp\Big[\, \Phi_R \d_{\a} {}^{\b} 
~+~ \frac 12 \Phi_{\un a \un b\, R} (\s^{\un a \un b})_{\a}^{~\b}
~+~ \frac 1{24} \Phi_{\un a \un b \un c \un d\, R} (\s^{\un a 
\un b \un c \un d})_{\a}^{~\b} \, \Big]
\end{equation}
where the 10D Pauli matrices above correspond to the right-handed
spacetime chiral projection of the 10D gamma matrices! This structure
introduces 256 (1,0) righton superfields ($\Phi_R$, $\Phi_{\un a \un b
\, R}$ and $\Phi_{\un a \un b \un c \un d\, R}$) instead of the
familiar 496 (1,0) righton superfields of the $E_8 \otimes E_8$
model.  The $\s$-model associated with this alternative construction 
is obtained by using the group elements in (2.9) to replace those in 
(2.4).  As well the quantity $\G_+$ below (2.7) must be replaced by
\begin{equation}
\G_+ ~\to ~ \Pi_+ {}^{\un a} \Big[ \,
(\nabla_{\un a} {\rm A} )~ 
\d_{\a} {}^{\b} ~+~ \frac 12 {\rm F}_{\un a \un b \un c}~ 
(\s^{\un b \un c})_{\a}^{~\b}  ~+~ \frac 1{24} {{\rm F}^{(+)}}_{\un a 
\un b  \un c \un d \un e}~  (\s^{\un b \un c \un d \un 
e})_{\a}^{~\b} \, \Big]
\end{equation}
The R-R bosonic fields here are precisely what is needed so that their
addition to the NS-NS fields describes the bosonic spectrum of 10D, 
type-IIB supergravity. The $\s$-model we have described can easily 
be studied upon performing dimensional compactification.  These
actions provide an intrinsic description of the models that were
recently studied by Maharana \cite{2B}.
 
Thus, we argued that among the family of 10D heterotic string theories,
by different choices of the matrix generators we could describe all
previous known models in a uniform manner.  As we found 
there seemed to exist one more member of this family that corresponds
to replacing the internal compact group generators by the non-compact
chiral 10D Pauli matrices and gives rise to a heterotic string model 
with a N = 2B supersymmetry.

Finally there is an intriguing interpretation that we can give to
(2.3) as modified in (2.9) and (2.10).  Within the context of superfield
theories, it is well known that it is possible to use low N superfields
to realize a theory with a higher than N supersymmetry.  Some examples
of this are the use of 4D, N = 1 superfield perturbation theory to 
describe 4D, N = 4 supersymmetric Yang-Mills theory \cite{GRS} or to
describe 4D, N = 2 supersymmetric Yang-Mills theory \cite{G}. In the
same way, our work suggests that it is possible to realize the N = 2B
superstring as a theory of N = 1 superstrings coupled to a certain
``matter superstring'' containing the R-R sector p-forms.

\sect{10D Green-Schwarz Heterotic Sigma-Models}

~~~~In the last section, we used (1,0) superfields to review the
situation of (1,0) supergravity heterotic $\s$-models.  Since it is 
our eventual goal to show that the last model above is very closely 
related to the type-B and type-O bosonic string theories, as a first 
step we begin to eliminate the (1,0) world sheet supersymmetry by 
going to a Green-Schwarz type formulation. The relevant starting points
are (2.5) and (2.7).   The action in (2.5) is replaced by ${\cal S}_{GS}$ 
$$
S_{GS} ~=~\int d^2 \s \,V^{-1}\Bigl [ - {\scriptstyle {\frac 12 }} \P_{\pp}^{~ 
\un a} \P_{\mm~\un a}~+~ \int_{0}^{1}dy \, \hat {\P}_{y}~^{\un C}\hat{\P}_{
\pp}^{~\un B}\hat{ \P}_{\mm}~^{\un A}\hat{G}_{\un A\un B\un C} ~+~
\Phi(Z) {\cal R}(V)  ~\Bigr ] ~~~, ~~~ $$
$$\P _{\pp}~^{\un A}~=~V_{\pp}^{ ~m}( \pa_{m}Z^{\un M}) E_{\un M}~^{\un 
A}~~,~~
\P _{\mm}~^{\un A}~=~V_{\mm}~^{m}(\pa_{m}Z^{\un M}) E_{\un M}~^{
\un A} ~~,$$
\begin{equation}
\hat{Z}^{\un M}~=~Z^{\un M}(\s,\t,y)~~,~~\hat{\P}_{y}{}^{\un 
A}~=~(\pa_{y}\hat{Z}^{\un M})E_{\un M} {}^{\un A} ({\hat Z})~~,~~
\hat{G}_{\un A\un B\un C} = G_{\un A\un B\un C}(\hat{Z})~.
\end{equation}
here $Z^{\un M}(\t, \s)$ is the superstring coordinate
($Z^{\un M}(\t, \s) \equiv (\Theta^{\m}(\t, \s), X^{\un m}(\t,\s))$,
$G_{\un A \un B\un C}$ is the field strength supertensor for a super
2-form $b_{\un A \un B}(Z)$, $\Phi(Z)$ is the spacetime dilaton and 
${\cal R}(V)$ is the world sheet curvature tensor.  Since only in the 
cases of the $E_8 \otimes E_8$ or $SO(32)$ theories is there spacetime 
supersymmetry\footnote{The only reason to use a Green-Schwarz formulation 
is precisely the presence of spacetime \newline ${~~~~~}$ supersymmetry.}, 
we do not at this stage have to worry about the introduction of a 
term to accommodate the tachyon that occurs for the other heterotic 
strings.

The replacement action for (2.7) can be obtained in the following
manner. The action in (2.7) is a (1,0) superfield action, which
using standard techniques, we have many times previously analyzed in
terms of its component fields. The component field formulation contains
fermions (none of which are dynamical), so to pass to an action
that is compatible with the Green-Schwarz type action above in (3.1), 
we simply set all fermions to zero. This leaves the action
\begin{equation}
\eqalign{
S_R =  - { 1 \over 4 \pi }
 \int d^2 \s \, &V^{-1} ~Tr \{~ (\cdp U^{-1} ) (\cdm U )
~+~ \l_{\mm} {}^{\pp} ( U^{-1} D_{\pp} U )^2  \cr
+~ & \int_{0}^1 dy (  {\tilde U}^{-1} { d~~ \over dy } {\tilde U})
[~ (\cdp {\tilde U}^{-1} ) (\cdm {\tilde U} ) \,-\,  (\cdm
{\tilde U}^{-1} ) (\cdp {\tilde U} ) ~ ] ~ \cr
- 2 & \, \P_{\pp}~^{\un B} \, \G_{\un B}^{~~\hat a}
t_{\hat a}  ( U^{-1} \cdm U ) ~  \} ~~ ,~~ } 
\end{equation}  
with $D_{\pp} U \equiv \cdp U - i \P_{\pp}~^{\un B} {\G}_{\un B}^{~~\hat a}
U t_{\hat a}$. The quantity $U \equiv exp [i {\phi}^{\hat a }_R (\t ,
\s) t_{\hat a }]$ is an element of an {\it {a}} {\it {priori}} arbitrary 
group.  The matrices $t_{\hat a}$ generate a compact Lie algebra for the
right-gauge group ${\cal G}_R$ where $\hat a = 1 , \dots ,d_G$, $ [ t_{
\hat a },~t_{\hat b } ] = i f_{\hat a \hat b}{}^{\hat c} t_{\hat c }$,
$f_{\hat a \hat b \hat c} f^{\hat a \hat b}{}_{\hat d} = c_2 {\d}_{\hat
c \hat d }$, and $ Tr (t_{\hat a}  t_{\hat b } ) = 2 k \d_{\hat a \hat
b}$.  Above, we have used the notation (${\cal D}_{\pp}, \, {\cal D}_{\mm}$) 
to denote the world-sheet two-dimensional gravitationally covariant 
derivative. To describe the two tachyon-free 10D theories, we pick 
${\cal G}_R = E_8 \otimes E_8$ or $SO(32)$, respectively.

At this stage, we once again replace the compact group generators and
their corresponding 2D righton fields as described in (2.9) but here
all the superfields (the $\Phi$'s) are replaced by a 2D
Duffin-Kemmer-Petiau field $\phi_{\a}{}^{\b}$
\begin{equation}
 \phi_{\a}{}^{\b} ~=~ \phi_R \d_{\a} {}^{\b} 
~+~ \frac 12 \phi_{\un a \un b\, R} (\s^{\un a \un b})_{\a}^{~\b}
~+~ \frac 1{24} \phi_{\un a \un b \un c \un d\, R} (\s^{\un a 
\un b \un c \un d})_{\a}^{~\b} ~~~.
\end{equation}
One final step is that the spacetime gauge superconnection and
generators on the last line of (3.2) must undergo the replacement
\begin{equation}
\G_{\un A}^{~~\hat a} t_{\hat a} ~\to ~ \Big[ \,
(\nabla_{\un A} {\rm A} )~ \d_{\a} {}^{\b} ~+~ \frac 12 {\rm F}_{\un A 
\un b \un c}~ (\s^{\un b \un c})_{\a}^{~\b}  ~+~ \frac 1{24} {{\rm 
F}^{(+)}}_{\un A \un b  \un c \un d \un e}~  (\s^{\un b \un c \un d 
\un e})_{\a}^{~\b} \, \Big] ~~~.
\end{equation}
Thus, we recover (within the context of a Green-Schwarz formulation)
the exact same result as seen from the NSR formulation. Namely, a 
heterotic 10D, N = 1 Green-Schwarz $\s$-model with manifest $E_8 
\otimes E_8$ symmetry is replaced by a model where the spectrum of 
field strength superfields in the R-R sector is exactly what one needs 
to describe the type-IIB theory.

\sect{On to Bosonic Type-B and Type-O Theories}

~~~~In this final section we complete the journey from the heterotic
type-IIB $\s$-model to the purely non-supersymmetric type-B and 
type-O models.  This is done in several stages.  First since these
final models are bosonic, we can begin with (3.1) and simply set
the Grassmann superstring coordinates ($\Theta$) identically to zero.
Since we want to allow coupling to the tachyon, we also add one additional
term to find the NS-NS sector of the type-B theory takes the form
\begin{equation}
\eqalign{
S_{NS}^B &=~ { 1 \over 2 \pi {\a}' } \int d^2 \s ~ V^{-1} [   \half 
( g_{ \un m \un n}(X) ~ + ~ b_{ \un m \un n}(X) ) ~ (\nabla_{\pp} 
{X}^{\un m}) (\nabla_{\mm} { X}^{\un n}) ~ ]  \cr
&{~~~~~~~}+~ \int  d^2 \s ~ V^{-1} \, [~ \Phi(X) {\cal R}  ~+~ T(X)  ~ ] \cr
&\equiv~ \int d^2 \s  V^{-1}
 [ \frac12 (\eta_{\un a \un b} + b_{\un a \un b}(X))
\P_{\pp} {}^{\un a} \P_{\mm} {}^{\un b}
~+~  \Phi(X) {\cal R} ~+~ T(X)  ~ ] ~~~. }
\end{equation}
This looks almost exactly like (2.5). There is one important
difference of course.  All of the quantities in (2.5) are (1,0) 
superfields whereas all quantities appearing here are ordinary 2D 
fields.

For the righton R-R sector of the type-B theory, we take the action 
of (3.2) and simply set to zero the Grassmann superstring coordinates
($\Theta$) to zero.
\begin{equation}
\eqalign{
S_{R-1}^B  =  - { 1 \over 4 \pi }
 \int d^2 \s \, &V^{-1} ~Tr \{~ (\cdp U_R^{-1} ) (\cdm U_R )
~+~ \l_{\mm} {}^{\pp} ( U_R^{-1} D_{\pp} U_R )^2  \cr
+~ \int_{0}^1 dy &(  {\tilde U_R}^{-1} { d~~ \over dy } {\tilde U_R})
[~ (\cdp {\tilde U_R}^{-1} ) (\cdm {\tilde U_R} ) \,-\,  (\cdm
{\tilde U_R}^{-1} ) (\cdp {\tilde U_R} ) ~ ] ~ \cr
- 2  \, \P_{\pp}~^{\un a}
&\Big[ \, (\nabla_{\un a} {\rm A}^1 ) ~+~ \frac 12 {\rm F}_{\un a \un b 
\un c}^1 ~ (\s^{\un b \un c})  ~+~ \frac 1{24} {{\rm F}^{(+)}}_{\un a 
\un b  \un c \un d \un e}~  (\s^{\un b \un c \un d \un e}) \, \Big] \,  
( U_{R}^{-1} \cdm U_R ) ~  \} ~~ ,~~ } 
\end{equation}  

The actions in (4.1) and (4.2) cannot comprise the entirety of the
type-B theory. We obtained the latter of these directly from the
GS action of the previous section. To complete the type-B theory
we must construct the ``mirror'' to (4.2) above. In the mirror action
all indices of the $\pp$ type are exchanged for the $\mm$ type and 
vice-versa.  However, this is not sufficient. In addition we must 
replace the non-supersymmetric limit of $U_{R-1}$ by its mirror also. 
This is done as follows.

In 10D, the notion of chiral spinors exists. In fact, the matrices that
appear in (2.9) are in such a basis. There also exist 10D matrices
where we exchange the handedness of the spacetime spinor indices
(i.e. $\a \to {\dot \a}$ etc.). Thus we may introduce $U_{L}$
according to the definition,
\begin{equation}
U_{L} ~\equiv~ exp\Big[\, \phi_L \d_{\dot \a} {}^{\dot \b} 
~+~ \frac 12 \phi_{\un a \un b\, L} (\s^{\un a \un b})_{\dot \a}^{~
\dot \b} ~+~ \frac 1{24} \phi_{\un a \un b \un c \un d\, L} (\s^{\un a 
\un b \un c \un d})_{\dot \a}^{~\dot \b} \, \Big] ~~~
\end{equation}
which introduces 256 lefton fields ($\phi_L$, $\phi_{\un a \un b
\, L}$  and $\phi_{\un a \un b \un c \un d\, L}$). Like its right
handed mirror counterpart, the argument of the exponential function
above can be shown to form a Lie algebra.  In fact, its composition
law is exactly the same as that given for the right handed one in
ref. \cite{Idea}, with the exception that the term involving the
Levi-Civita tensor has the opposite sign.  The complete ``mirror'' 
action to (4.2) is given by 
\begin{equation}
\eqalign{
S_{R-2}^B =  - { 1 \over 4 \pi }
 \int d^2 \s \, &V^{-1} ~Tr \{~ (\cdp U_L^{-1} ) (\cdm U_L )
~+~ \l_{\pp} {}^{\mm} ( U_L^{-1} D_{\mm} U_L )^2  \cr
-~ \int_{0}^1 dy &(  {\tilde U_L}^{-1} { d~~ \over dy } {\tilde U_L})
[~ (\cdp {\tilde U_L}^{-1} ) (\cdm {\tilde U_L} ) \,-\,  (\cdm
{\tilde U_L}^{-1} ) (\cdp {\tilde U_L} ) ~ ] ~ \cr
- 2  \, \P_{\pp}~^{\un a}
&\Big[ \, (\nabla_{\un a} {\rm A}^2 ) ~+~ \frac 12 {{\rm F}}_{\un a 
\un b \un c}^2 ~ ({\Bar \s}^{\un b \un c})  ~+~ \frac 1{24} {{\rm 
F}^{(-)}}_{\un a \un b  \un c \un d \un e}~  ({\Bar \s}^{\un b \un c 
\un d \un e}) \, \Big] \,  ( U_{L}^{-1} \cdm U_L ) ~  \} ~~ .~~ } 
\end{equation}  
On the final line above, we have placed bars above the 10D Pauli matrices
to indicate that these are the ones with the dotted spacetime spinor
indices. So the complete non-linear $\s$-model for the type-B
theory is just
\begin{equation}
S_{Tot}^B ~=~ S_{NS}^B~+~ S_{R-1}^B~+~S_{R-2}^B ~~~.
\end{equation}
It can be seen that the complete spectrum of spacetime fields includes
the graviton ($g_{ \un m \un n}$), axion ($b_{ \un m \un n} $), dilaton 
($\Phi$), tachyon ($T$),  two scalars (${\rm A}^1$ and ${\rm A}^2$
)\footnote{An interesting feature to note is that the scalars $A^1$ and $A^2$
possess ``moduli'' (i.e. we may \newline ${~~~~~}$ shift them by constants 
without affecting the action) unlike the scalars that usually appear 
\newline ${~~~~~}$ in Lefton-Righton Thirring Models \cite{BDG}.}, two 
Kalb-Ramond fields (${\rm A}_{ \un m \un n}^1$ and ${\rm A}_{\un m \un 
n}^2$), 4-form ${\rm A}_{ \un a \un b \un c \un d}^+$ with a self-dual 
field strength ${{\rm F}^{(+)}}_{\un a \un b \un c \un d \un e}$ and a 
second 4-form ${\rm A}_{\un a \un b \un c \un d}^-$ with an anti-self-dual 
field strength ${{\rm F}^{(-)}}_{\un a \un b \un c \un d \un e}$.

The key to our successful construction of the model described 
immediately above is the suggestion that currents associated with 
internal symmetries can sometimes be traded for currents associated with
the Clifford algebra of the spacetime spinors. We first did this in
the 10D supersymmetric theory.  If this conjecture is accepted, then 
the reverse is likely to also be true.  With this as a working assumption,
we will now show that the type-O string \cite{Sag} propagating in
the presence of its tachyon and massless modes can be constructed
from the corresponding type-B theory.

So we once more will trade some of the currents.  In particular
we note that by keeping only the middle terms in the exponential
of (3.3) and (4.3) we also maintain the structure of a Lie group.
So the simple idea is to trade the currents associated with the
0-forms and 4-forms and replace them by currents associated with
$SO(32)$. This is all easily done for the $\s$-models and leads
to the following type-O action.  The final answer is rather
tedious to write only because it possesses a number of different R-R
sectors but takes the form
\begin{equation}
S_{Tot.}^{O} ~\equiv ~ S_{NS}^{O} ~+~ S_{R-1}^O ~+~ S_{R-2}^O
~+~ S_{R-3}^O ~+~ S_{R-4}^O ~~~.
\end{equation}
Below we give each sub-action.
\begin{equation}
\eqalign{
S_{NS}^{O} &=~ \int d^2 \s  V^{-1} [ \frac12 \,\eta_{\un a \un b}
\P_{\pp} {}^{\un a} \P_{\mm} {}^{\un b} ~+~  \Phi(X) {\cal R} 
~+~ T(X)  ~ ] ~~~. }
\end{equation}

\begin{equation}
\eqalign{
S_{R-1}^O  =  - { 1 \over 4 \pi }
 \int d^2 \s \, &V^{-1} ~Tr \{~ (\cdp U_{R}^{-1} ) (\cdm U_{R} )
~+~ \l_{\mm} {}^{\pp} ( U_{R}^{-1} D_{\pp} U_{R} )^2  \cr
+~ \int_{0}^1 dy &(  {\tilde U_{R}}^{-1} { d~~ \over dy } {\tilde 
U_{R}})
[~ (\cdp {\tilde U_{R}}^{-1} ) (\cdm {\tilde U_{R}} ) \,-\,  (\cdm
{\tilde U_{R}}^{-1} ) (\cdp {\tilde U_{R}} ) ~ ] ~ \cr
-   \, \P_{\pp}~^{\un a} &\Big[ \,  {\rm F}_{\un a \un b \un c}^1 ~ 
(\s^{\un b \un c})  \, \Big] \,  ( U_{R}^{-1} \cdm U_{R} ) ~  
\} ~~ ,~~ } 
\end{equation}

\begin{equation}
\eqalign{
S_{R-2}^O  =  - { 1 \over 4 \pi }
 \int d^2 \s \, &V^{-1} ~Tr \{~ (\cdp U_{L}^{-1} ) (\cdm U_{L} )
~+~ \l_{\mm} {}^{\pp} ( U_{L}^{-1} D_{\pp} U_{L} )^2  \cr
-~ \int_{0}^1 dy &(  {\tilde U_{L}}^{-1} { d~~ \over dy } {\tilde 
U_{L}})
[~ (\cdp {\tilde U_{L}}^{-1} ) (\cdm {\tilde U_{L}} ) \,-\,  (\cdm
{\tilde U_{L}}^{-1} ) (\cdp {\tilde U_{L}} ) ~ ] ~ \cr
-  \, \P_{\mm}~^{\un a} &\Big[ \,{\rm F}_{\un a \un b \un c}^2 ~ 
({\bar \s}^{\un b \un c}) \, \Big] \,  ( U_{L}^{-1} \cdp U_{L} ) ~ 
 \} ~~ ,~~ } 
\end{equation}

\begin{equation}
\eqalign{
S_{R-3}^O ~\equiv ~  \int d^2 \s \, &V^{-1} \Big[ \, i \, \psi_- (
\, \cdp ~+~  i \P_{\pp}~^{\un a} \,  A_{\un a}^{\hat a} ~t_{\hat a}
 \, )  \psi_- ~ \Big] ~~~,~~{~~~~~~~~~~~~~~~~~~~~} } 
\end{equation}

\begin{equation}
\eqalign{
S_{R-4}^O ~\equiv ~  \int d^2 \s \, &V^{-1} \Big[ \, i \, \psi_+ (
\,  \cdm ~+~ i \P_{\mm}~^{\un a} \,  {\tilde A}_{\un a}^{\hat a} 
~t_{\hat a}  \, )  \psi_+ ~ \Big] 
~~~.~~ {~~~~~~~~~~~~~~~~~~~~} } 
\end{equation}  
In the latter two equations above, we have introduced 64 right-moving
and 64 left-moving Majorana fermions ($\psi_-$ and $\psi_+$, respectively) 
on the worldsheet.  These are the degrees of freedom that were traded by 
removing the lefton and righton 0-forms and 4-forms in $U_R$ and $U_L$ in 
the type-B theory.

The complete spectrum of spacetime fields includes the graviton 
($g_{ \un m \un n}$), dilaton ($\Phi$), tachyon ($T$), two sets 
of $SO(32)$ gauge fields ($A_{\un a}^{\hat a}$ and $ {\tilde A}_{\un 
a}^{\hat a}$) and two Kalb-Ramond fields (${\rm A}_{\un m \un n}^1$ 
and ${\rm A}_{\un m \un n}^2$). This is the spectrum of the type-O
bosonic string at low orders.

\noindent
{\bf {Acknowledgment; }} \newline \noindent
${~~~~}$One of the authors (SJG) wishes to acknowledge M. T. Grisaru,
M. Gaberdiel, Nikita Nekrasov and B. Zweibach and for useful discussions.  
Additional thanks go to R. Whittlesay and N. Buckner who acted as 
the catalysts for these discussions.

\newpage 
\noindent{{\bf {APPENDIX A: Kalb-Ramond Matter Field and 
4D String $\s$-models  }}}

One interesting consequence of the construction of the model in
ref. \cite{Idea} is that a compactification of the model reveals
how 4D Kalb-Ramond matter fields can be coupled to the world
sheet in NSR or GS string $\s$-models.  The idea is that the
previous work utilizing Lefton-Righton Thirring Models \cite{BDG,LRTM} 
can appropriately be modified to include Kalb-Ramond fields as
long as the Kalb-Ramond fields do not carry any internal
charges. In this brief appendix we will demonstrate how this
construction is carried out within the confines of the 4D, N = 8
(1,0) supergravity NSR $\s$-model. We only pick this choice
because of its simplicity. Any model with smaller values of N 
(or even larger values of D) can be treated by the same techniques
as those we use below.

The 4D, N = 8 (1,0) supergravity NSR $\s$-model is described by
using the NSR sector in (2.5) but the R-R sector is replaced by
$$
\eqalign{
S_{R-5} = \int d^2 \s d\z^- E^{-1} i\frac12 [~
  ( &L_{=}^{\hat \a} + \G_{=}^{\hat \a} ) ( L_+^{\hat\a}
           - \L_+ {}^{=} (L_{=}^{\hat\a} + \G_{=}^{\hat\a}))
         +   L_+^{\hat\a} \G_{=}^{\hat\a} \cr
+ &  ( R_+^{\hat I} + 2 \G_+^{\hat I}) R_{=}^{\hat I}
        - i \L_{=}{}^{\pp} (R_+^{\hat I} + \G_+^{\hat I} )
         \de_+ (R_+^{\hat I}+\G_+^{\hat I}) \cr
+ & 4 S^{\a \hat I} \Sc R_{=}^{\hat I}  \Sc L_+ ^{\hat \a}
 - ~ 4  \L_+{}^{=} (M^{-1})^{\hat I\hat K}\Phi^{\hat \a \hat I}
        \Phi^{\hat \a \hat J} \S_{=}^{\hat J} \S_{=}^{\hat K} \cr
- & 4 i \L_{=}{}^{\dpx} \Phi^{\hat \a \hat I} \Sc L_+^{\hat \a} \nabla_+
    ( \Phi^{\hat \b \hat I} \Sc L _+^{\hat \b} )~]  ~~~, \cr}
\eqno(A.1)
$$

$$
\eqalign{
\Sc L _+ ^{\hat \a} = &\  L _+ ^{\hat \a} - \L _+{}^{=} ( L_{=}^{\hat \a}
        +\G_{=}^{\hat\a}) ~~, ~~ L_{A}^{\hat \a} \equiv \de_{A}
        \varphi_L {}^{\hat \a}  ~~, ~~ R_{A}^{\hat I} \equiv
        \de_{A} \varphi_R {}^{\hat I}  ~~, \cr
\Sc R _{=} ^{\hat I} = &\ R _{=} ^{\hat I} - i [\L _{=}{}^{\dpx}
        \de _+ (R_+ ^{\hat I} + \G_+^{\hat I}) + \ha (\de _+ \L_{=}{}^{\dpx})
        (R_+ ^{\hat I} + \G _+ ^{\hat I}) ] ~~, \cr
        \S_{=}^{\hat I} = &\ \Sc R_{=}^{\hat I} -2 i [ \L_{=}{}^{\dpx}
        \nabla_+ (\Phi^{\hat \b \hat I} \Sc L_+^{\hat \b} )+\ha (\nabla_+ 
        \L_{=}{}^{\dpx} ) \Phi^{\hat \b \hat I}
        \Sc L_+^{\hat \b} ] ~~, \cr
~~ ( M )^{\hat I \hat J} = & \ \d^{\hat I \hat J} -4 i ( \nabla_+
        \L _+{}^{=})  \L_{=}{}^{\dpx} \Phi^{\hat
        \a  \hat I} \Phi^{\hat \a \hat J} ~~ ,} \eqno(A.2)
$$

$$
\eqalign{
\G_{=}{}^{\hat \a} & \equiv \P_{=}{}^{\un a} A_{\un a}{}^{\hat \a}(X)
~~,  ~~ A_{\un a}{}^{\hat \a} = ({\widetilde A}_{\un a}^{[ij]})~~, \cr
\G_+{}^{\hat I} & \equiv  \P_{+} {}^{\un a} A_{\un a}{}^{\hat I}
(X)~~, ~~
A_{\un a}{}^{\hat I} = (A_{ \un a}, A_{\un a}^{[i^{\prime}j^{\prime}]},
A_{ \un a}^{[i^{\prime}j^{\prime}] [ k^{\prime } l^{\prime}  ]} ) ~~.
\cr }  \eqno(A.3)
$$

$$
\eqalign{
 \F_{\hat \a \hat I}  & \equiv  ~ ( ~ \Phi_{[ij]} ,~
\Phi_{[ij] [i^{\prime}j^{\prime}]} ,~ \Phi_{[p^{ \prime}   
q^{\prime}]  [ i^{\prime} j^{\prime}] [ k^{\prime} l^{\prime}]}
,~ \d_{ i^{\prime} [i } \d_{j ]  j^{\prime} }
{\widetilde \Phi}_{[ k^{\prime}  l^{\prime}]} -
\d_{ k^{\prime} [i } \d_{j ]  l^{\prime} }
{\widetilde \Phi}_{[ i^{\prime}  j^{\prime}]} ~)  ~~~. }
\eqno(A.4)
$$
We see in addition to the graviton, axion and dilaton which
appear in the NSR sector, there also appear 28 spin-1 fields
$({\widetilde A}_{\un a}^{[ij]}, \, A_{ \un a}, A_{\un a}^{[i^{\prime}
j^{\prime}]}, \, A_{ \un a}^{[i^{\prime}j^{\prime}] [ k^{\prime } 
l^{\prime}  ]} )$ and 68 scalar fields $(\Phi_{[ij]} , \,
\Phi_{[ij] [i^{\prime}j^{\prime}]} ,~ \Phi_{[p^{ \prime}   
q^{\prime}]  [ i^{\prime} j^{\prime}] [ k^{\prime} l^{\prime}]}
, \, {\widetilde \Phi}_{[ k^{\prime}  l^{\prime}]} )$ which
complete the bosonic spectrum to that of 4D, N = 8 supergravity.

We now wish to carry out a duality transformation on the world 
sheet of the (1,0) string whereby some of the scalar fields
are replaced by Kalb-Ramond fields.  We will use the scalar
field ${\widetilde \Phi}_{[ k^{\prime}  l^{\prime}]}$ for the 
purpose of illustration.

We first set to zero both ${\widetilde \Phi}_{[ k^{\prime} l^{\prime}]}$
and $\phi_R^{[ k^{\prime} l^{\prime}]}$.  The former operation
eliminates the spacetime scalar ${\widetilde \Phi}_{[ k^{\prime}  
l^{\prime}]}$ from among the background fields and the latter
operation eliminates the modes of the string which describe this
state. At this intermediate point, the theory will be inconsistent.
In order to restore the consistency, new modes on the world
sheet must be introduced. We can do this by introducing a new
righton WZNW model precisely of the form of the R-R terms in
(2.3).  However, the group element that corresponds to (2.4)
is here described by
$$
U_{R-2} ~\equiv~ exp\Big[\, \frac 12 \Phi_{\un a \un b\, R}^{[ k^{\prime} 
l^{\prime}]} ~(\s^{\un a \un b}) \otimes {\cal T}_{[ k^{\prime} 
l^{\prime}]} \, \Big]
\eqno(A.5)
$$
where now $\s^{\un a \un b}$ denotes the right-handed Pauli matrix 
Lorentz generator for 4D spacetime.  Also in the above expression
${\cal T}_{[ k^{\prime} l^{\prime}]}$ denotes a matrix representation
of a $U(1)^6$ group.  Note that the fact that this is an abelian  
group is absolutely critical.  The matrices $\s^{\un a \un b} \otimes
{\cal T}_{[ k^{\prime} l^{\prime}]}$ clearly form an algebra.
This would not necessarily be the case if ${\cal T}_{[ k^{\prime} 
l^{\prime}]}$ represented some non-abelian group.  So the duality on 
the worldsheet correspond to
$$
{\widetilde \Phi}_{[ k^{\prime} l^{\prime}]} ~ \to ~
\Phi_{\un a \un b\, R}^{[ k^{\prime} l^{\prime}]} ~~~.
\eqno(A.6)
$$
Finally to complete the introduction of the 4D spacetime Kalb-Ramond
field, we replace the $\G_+$ in (2.7) by
$$
\G_+ ~\to ~ \Pi_+ {}^{\un a} \, F_{\un a \, \un b \, \un c}^{[ k^{\prime} 
l^{\prime}]} ~(\s^{\un b \un c}) \otimes {\cal T}_{[ k^{\prime} 
l^{\prime}]} ~~~.
\eqno(A.7)
$$
We note that interestingly enough, the reduction of the type-IIA
theory is expected to possess precisely six matter Kalb-Ramond
fields.  Thus, the final model that we have discussed (or possibly
its mirror with respect to the duality transformation described above) 
might present an intrinsic approach to investigating heterotic
type-IIA duality totally within the confines of heterotic
string theory.


\end{document}



\begin{thebibliography}{66}

\bibitem{Idea}S. James Gates, Jr. and V.G.J. Rodgers, Phys. Lett. 
{\bf {357B}} (1995) 552.

\bibitem{HET}D.~Gross, J.~Harvey, E.~Martinec and R.~Rohm, Phys.~Lett. 
{\bf {54}} (1985) 502; idem. Nucl.~Phys. {\bf {B256}} (1985) 253; idem. 
Nucl.~Phys. {\bf {B267}} (1986) 75.

\bibitem{Typ2BSG}M.~T.~Grisaru, P.~Howe, L.~Mezincescu, B.~Nilsson and 
P.~Townsend, Phys. Lett. {\bf {162B}} (1985) 116;  M.~T.~Grisaru, H.~Nishino 
and D.~Zanon, Phys. Lett. {\bf {206B}} (1988) 625; S. Bellucci, S.~J.~Gates, 
Jr., B. Radak, P.~Majumdar and S. Vashakidze, Mod. Phys. Lett. {\bf {A21}} 
(1989) 1985.

\bibitem{GS2}M.~B.~Green and J.~H.~Schwarz, Nucl. Phys. {\bf {B243}} 
(1984) 285.

\bibitem{GN}S.~J.~Gates, Jr. and H.~Nishino, Class. \& Quant.~Grav. 
{\bf {8}} (1991) 809.

\bibitem{Nil}Martin Cederwall, Alexander von Gussich, Bengt E. W. Nilsson, 
Per Sundell, Anders Westerberg, ``Dirichlet Super-p-Branes in Ten-Dimensional 
Type IIA and IIB Supergravity,'' hep-th/9611159; idem. ``The Dirichlet 
Super-Three-Brane in Ten-Dimensional Type IIB Supergravity,'' 
hep-th/9610148.

\bibitem{Sieg}W.Siegel, Nucl.~Phys. {\bf {B238}} (1984) 307.

\bibitem{BGM}R. Brooks, S. J. Gates, Jr. and F. Muhammad, Nucl.~Phys. 
{\bf {B268}} (1986) 599.

\bibitem{GaSg}S.~J.~Gates, Jr. and W.~Siegel, Phys.~Lett. {\bf {206B}} 
(1988) 631.

\bibitem{het10s}L. J. Dixon and J. A. Harvey, Nucl. Phys. {\bf {B274}} 
(1986) 93; L. Alvarez-Gaum\' e, P. Ginsparg, G. Moore and C. Vafa,
Phys.~Lett. {\bf {171B}} (1986) 155; H. Kawai, D.C. Lewellen, S.-H.H.
Tye, Phys. Rev. {\bf {D34}} (1986) 3794; P. Forg\' acs, Z. Horv\' ath, 
L. Palla and P. Vecserny\' es, Nucl. Phys. {\bf {B308}} (1988) 477.

\bibitem{cups}D. Friedan, Phys.~Rev.~Lett. {\bf {45}} (1980) 1057;
S. J. Gates, Jr. C. M. Hull and M. Ro\v cek, Nucl. Phys. {\bf {B248}} 
(1984) 157; E..~S.~Fradkin and A.~A.~Tseytlin, Phys.~Lett. {\bf {158B}} 
(1985) 316; idem. Nucl. Phys. {\bf {B261}} 
(1985) 1.

\bibitem{SEN}A. Sen, Phys. Rev. {\bf {32D}} (1985) 2162; Phys. Rev.
 Lett. {\bf {55}} (1985) 1846. 

\bibitem{BDG}D. Depireux, S. J. Gates, Jr. and S. Bellucci, Phys.~Lett. 
{\bf {232B}}  (1989) 67.

\bibitem{2B}J. Maharana, ``S-Duality and Compactification of Type IIB
Superstring Action,'' Fermilab preprint, FERMILAB-PUB-97/046-T,
hep-th/9703009.

\bibitem{GRS}M.~T.~Grisaru, M.~Ro\v cek and W.~Siegel, Phys.~Rev.~Lett.
{\bf {45}} (1980) 1063; idem. Nucl.~Phys. {\bf {B183}} (1981) 141.

\bibitem{G}S. J. Gates, Jr., Nucl.~Phys. {\bf {B238}} (1984) 349.

\bibitem{OST}O. Bergman and M. Gaberdiel, Harvard Preprint HUTP-97/A003,
hep-th/9701137, (Jan., 1997).

\bibitem{Sag}A. Sagnotti, {\it {Open Strings and Their Symmetry Groups}},
Carg\' ese '87 , ``Nonperturbative Quantum Field Theory,'' eds. G. Mack 
{\it {et}}. {\it {al}}. (Pergammon Press, 1988); A. Sagnotti
and Massimo Bianchi, Phys. Lett. {\bf {247B}} (1990) 517.
.

\bibitem{LRTM}D. Depireux, S. J. Gates, Jr. and Q-Han Park,  Phys. Lett. 
{\bf {224B}} (1989) 364.



\end{thebibliography}
\end{document}